\documentclass[]{emulateapj}
\usepackage{url}

\def\gtsima{$\; \buildrel > \over \sim \;$}
\def\ltsima{$\; \buildrel < \over \sim \;$}
\def\gsim{\lower.5ex\hbox{\gtsima}}
\def\lsim{\lower.5ex\hbox{\ltsima}}
\def\simgt{\lower.5ex\hbox{\gtsima}}
\def\simlt{\lower.5ex\hbox{\ltsima}}
\def\simpr{\lower.5ex\hbox{\prosima}}
\def\la{\lsim}
\def\ga{\gsim}

\def\eg{{\frenchspacing\it e.g. }}
\newcommand{\be}{\begin{equation}}
\newcommand{\ba}{\begin{eqnarray}}
\newcommand{\ee}{\end{equation}}
\newcommand{\ea}{\end{eqnarray}}

\begin{document}



\title{The 3D skeleton of the SDSS}

\author{    Thierry\,Sousbie$^1$,    Christophe
Pichon$^{1,2}$,  H\'el\`ene\,Courtois$^1$, St\'ephane Colombi$^{2}$, Dmitri Novikov$^{3}$} \affil{\footnotesize $^1$ CRAL,
Observatoire de Lyon  ,   69561   Saint   Genis   Laval   Cedex,   France;
sousbie@obs.univ-lyon1.fr}    \affil{\footnotesize    $^2$    Institut
d'Astrophysique  de  Paris, UMR 7595, UPMC, 98  bis  boulevard  d'Arago, 75014  Paris,
France} \affil{\footnotesize $^3$ Astrophysics Group, Imperial College, Blackett Laboratory, 
Prince Consort Road, London, SW7 2AZ, UK}

\begin{abstract}
The length of the three-dimensional filaments observed in the fourth
public data-release of the SDSS is measured using the {\em local
skeleton} method. It consists in defining the set of points where the gradient of the
smoothed density field is extremal along its isocontours,
with some additional constraints on local curvature to probe actual ridges in the galaxy
distribution. A good fit to the mean filament length per unit volume, $\cal{L}$, 
in the SDSS survey is found to be  
${\cal{L}}=(52500\pm6500)\, (L/{\rm Mpc})^{-1.75\pm0.06}\rm{Mpc}/(100\,\rm{Mpc})^{3}$
for $8.2 \leq L \leq 16.4$ Mpc, where $L$ is the smoothing length in Mpc. This result, which
deviates only slightly, as expected, from the trivial behavior ${\cal{L}} \propto L^{-2}$,
is in excellent agreement with a $\Lambda$CDM cosmology, as long as
 the matter density 
parameter remains in the range $0.25 < \Omega_{\rm matter} < 0.4$ at one sigma confidence level, considering the universe is flat. 
These measurements, which are in fact dominated by linear dynamics, 
are not significantly sensitive to observational biases such as redshift 
distortion, edge effects, 
incompleteness, and biasing between the galaxy distribution and the dark matter
distribution. Hence it is argued that the local skeleton is a rather promising and discriminating tool
for the analysis of filamentary structures in three-dimensional galaxy surveys.
\end{abstract}

\keywords{methods: data analysis, statistical --- cosmology: large-scale structure }

\section{Introduction}
\label{sec:intro}

\begin{figure}[t]
\begin{minipage}[b]{0.95 \linewidth}
\centering \includegraphics[height=6cm,width=8cm]{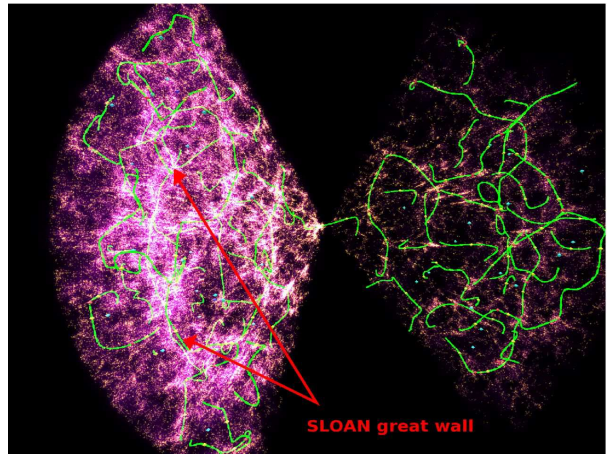}

\caption{Derived 3D skeleton in a slice of the SDSS for a smoothing
length $L=16.4$ Mpc. This animation still frame shows how well
structures are captured, notably the SLOAN great wall. The mpeg movie
version is available at \texttt{http://www.projet-horizon.fr/article173.html}.\label{fig1}}
\end{minipage}
\end{figure}

From the Great Wall of CFA1  \citep[]{Geller89} to the very long
filaments seen in the SDSS  \citep[]{Gott05} and the 2DF
\citep[]{Colless01},  the ever growing size of the largest structures
observed in the three-dimensional galaxy distribution has remained  a
challenge to models of large scale structure formation.  It is
therefore of prime importance to find a robust way to identify
filaments in the Universe and to characterize them, \eg through their
length, thickness and/or average density. To achieve that, one usually
relies on the analysis of the morphological properties --\eg through structure
functions \citep[]{Babul92}, Minkowski functionals \citep[]{Mecke94},
shape finders \citep[]{Sahni98}-- of an excursion's connected components in overdense  regions of the catalog, which can be obtained
using friend-of-friend algorithms \citep[]{Zel82}, the minimum
spanning tree technique \citep[]{Barrow85,Dor03} or percolation on a
grid where the density field has been smoothly interpolated
\citep[]{Gott86,Dominik92}.

This letter works instead in the framework of Morse theory \citep[]{Colombi00},
and uses the approach proposed recently by \citeauthor{Nov06} (2006, hereafter NCD) 
and \citeauthor{SouSk06} (2006, hereafter SPCN), where
filaments are seen as a set of special field lines, 
departing from saddle points and converging
to local maxima while following the gradient of the density field, 
$\nabla \rho\equiv \partial \rho/\partial x_i\equiv \rho_i$.
However, the {\em skeleton} thus defined remains non-local,
which makes analytical calculations challenging and edge effects
difficult to cope with in real catalogs. To solve these issues, a local approximation of 
the skeleton was proposed by NCD in the 2D case, and generalized to 3D
by SPCN. Given the Hessian, ${\cal H}\equiv \partial^2 \rho/\partial x_i \partial x_j \equiv \rho_{ij}$, and its
eigenvalues, $\lambda_i$, ranked in decreasing order, the {\em local skeleton}
is defined as the set of points where ${\cal H}\cdot \nabla\rho =\lambda_1 \nabla\rho$
and $\lambda_2, \lambda_3 < 0$, to ensure that ridges of the density field
are probed. 

In this paper, the local skeleton is extracted from the SDSS DR4 galaxy catalog. Its total length per unit volume is measured and compared to that obtained in  
$\Lambda$CDM cosmologies. Relying on realistic mock catalogs, various 
effects such as incompleteness, survey geometry, cosmic variance, redshift distortions,
biasing and non-linear dynamics, are extensively tested. 

\begin{deluxetable*}{ccccccccc}
\tablecaption{Length per unit volume of the skeleton for different SDSS and mock samples
\label{tab1}}   
\tablewidth{0cm}   \tablehead{   \colhead{  }   &
\colhead{L} & \colhead{DR4-350} & \colhead{DR4-{VL350}} & \colhead{MOCK} &
\colhead{MOCK-{PS}}& \colhead{MOCK-AS}&\colhead{MOCK-NB} } \startdata
length density                        &16.4 & 372 & 363 & 390  $\pm$ 19&   395$\pm$ 16 & 362  &403 $\pm$ 17 \\ 
$[\rm{Mpc}/(100\,\rm{Mpc})^{3}]$     &10.9 & 795 & 772 & 796  $\pm$ 18&   815$\pm$ 19 & 740  &790 $\pm$ 24 \\ 
                                      &8.2  & 1271& 1299& 1272 $\pm$ 25&  1308$\pm$ 27 & 1285 &1204 $\pm$ 25 \\
\enddata
\end{deluxetable*}
\section{Observational and mock data samples}
\label{sec:thesamples}
A complete description of the Fourth Data Release of the Sloan Digital
Sky  Survey (DR4  SDSS) can  be  found in  \citeauthor{Ade05} (2006).  The  main
sample  used in  this paper  is extracted  from the  Catalog Archive
Server facility.   To  ensure  proper spectral identification of galaxies,
objects  with {\em specclass} $=2$ and {\em zconf} $>0.35$  are selected  in  the {\em specphoto} table. This yields a main sample containing $459,408$  galaxies. The completeness
in  apparent  magnitude  was  investigated and  is  achieved  for
$U_{\rm{SDSS}} < 19$.   Two  subsamples were extracted: a sample
cut at distance $d < 350$ Mpc containing $148,012$ galaxies (hereafter DR4-350),
and  a homogeneous,  volume-limited sample (hereafter DR4-VL350)  containing  $25,843$  galaxies  
selected  on  the basis of their  absolute magnitude $M_{\rm absU} < -17$ and $d < 350 $ Mpc.

To test the robustness of the measurements and compare observational
results to theoretical predictions, 
a large $\Lambda$CDM simulation was performed, using
the publicly available treecode GADGET-2 \citep[]{gadget2}, involving $512^3$
particles in a $1024\,h^{-1}\rm{Mpc}$ box,
and with the following cosmological parameters: $H_{0}=70$ km/s/Mpc, $\Omega_{\rm baryons}=0.05$, 
$\Omega_{\rm matter}=0.3$, $\Omega_{\Lambda}=0.7$ and normalization $\sigma_{8}=0.92$.
Various mock catalogs were extracted from this simulation, using MoLUSC \citep[]{SouMo06}. 
This tool is designed to build realistic mock galaxy catalogs from 
dark matter simulations of large volume but poor mass resolution, by
reprojecting, as functions of local phase-space density, 
the statistical properties of the galaxy distribution  
(type, spectral features, number density, etc.)
derived from semi-analytic models applied to
simulations of higher mass resolution.
Here, the results calculated by GalICS \citep[]{galics} on
a treecode simulation with $256^3$ particles in a cube of $150$ Mpc on a side are used
as inputs of MoLUSC.
According to the analyses of Blaizot et al. (2006), this simulation should provide
sufficient mass resolution to describe realistically the statistical properties of SDSS galaxies 
with $U_{\rm{SDSS}} < 19$. The advantage of using
MoLUSC is that it allows one to probe a realistic volume of the Universe without worrying about finite
volume or replication effects in the realization of mock catalogs themselves \citep[]{Blai05}. 
  
For the purpose of testing the skeleton properties, 
three different kinds of mock catalogs were built, all cut at a distance of $350$ Mpc: (1) the main
catalog is called MOCK and attempts to reproduce all the characteristics of
DR4-350 (redshift space distortion, incompleteness, survey geometry, etc.); (2)
MOCK-PS is identical to MOCK but uses the exact positions of the galaxies  
to test the effect of redshift distortion; (3)
MOCK-AS is an all-sky version of MOCK 
aimed to test the influence of survey geometry and finally (4) MOCK-NB is identical to MOCK but with dark matter particles (without density biasing). 
The volume of our simulation is approximately 30 times that covered by 
DR4-350, which  yields
an error bar reflecting cosmic variance from the dispersion
among 25 random realizations of MOCK. Note finally that measurements were also performed  directly on the
dark matter distribution simulation boxes and on the initial conditions of the simulation, to test the effects of nonlinear clustering.
\section{The 3D Skeleton: algorithm}
The details of the algorithm used to draw the local skeleton defined in
\S~\ref{sec:intro} are given in SPCN, so only a brief sketch
of it is given here:

(i) {\em interpolation and smoothing:} the first step consists in
performing cloud-in-cell interpolation \citep[]{Hock81} on a $512^3$
grid covering a $700$ Mpc cube embedding the survey.  To avoid
extra degeneracies while drawing the skeleton, the empty regions of
the cube are filled with a random distribution of galaxies with $1,000$
times smaller average density than inside the survey.  At the end of
the process, only the parts of the skeleton belonging to the original
survey are kept.  To warrant sufficient differentiability, convolution
with a Gaussian window of size $L$ is performed prior to computing the gradient and the Hessian using a finite difference
method. As argued in NCD, in order to avoid contamination by the grid,
discreteness and finite volume effects, respectively, the smoothing
scale should verify $L/\Delta \ga 4$, $L/\lambda \ga 1$ and $L/V^{1/3}
\la 20$ where $\Delta$ is the grid step, $\lambda$ the mean
interparticle distance and $V$ is the survey volume. As a result, the
following conservative scale range, $8.2 \leq L \leq 16.4$ Mpc, will be used for
the measurements performed in this paper.  

(ii) {\em surface intersection modeling:}
the second step of the algorithm consists in drawing the skeleton, noting
that it is embedded in the set of points verifying ${\cal H} \cdot \nabla \rho \times \nabla \rho=0$.
This leads to 3 conditions, $S_i(x_1,x_2,x_3)=0$, 
that define 3 surfaces intersecting along a common line.  The actual
method used to compute the surfaces $S_i=0$ as an assemblage of triangles and their intersections as a set
of connected lines relies on the classical marching cube algorithm \citep[]{Lor87}, 
as detailed further in SPCN. Additional conditions, namely that
the gradient should be aligned with the major axis of curvature -- in practice, 
$|\nabla \rho.{\bf u}_1| > \rm{max}\left(|\nabla \rho.{\bf u}_2|,|\nabla \rho.{\bf u}_3|\right)$ 
where ${\bf u}_i$ are the eigenvectors of ${\cal H}$-- and $\lambda_2,\lambda_3 < 0$ are enforced
locally after diagonalizing the Hessian. 

(iii) {\em cleaning:} some additional treatment has to be performed in regions where
the field becomes degenerate (\eg in the vicinity of critical points, 
$\nabla \rho=0$), as explained in detail in SPCN.
Finally, the  parts of the local skeleton which do not 
pass through any critical point are removed. As argued in SPCN, these parts are mostly irrelevant as they do not, 
in general, correspond to real filaments. 

\section{
Measurements and robustness vs observational biases}
\label{sec:length}

\begin{figure*}[t]
\begin{center} \includegraphics[width= \linewidth]{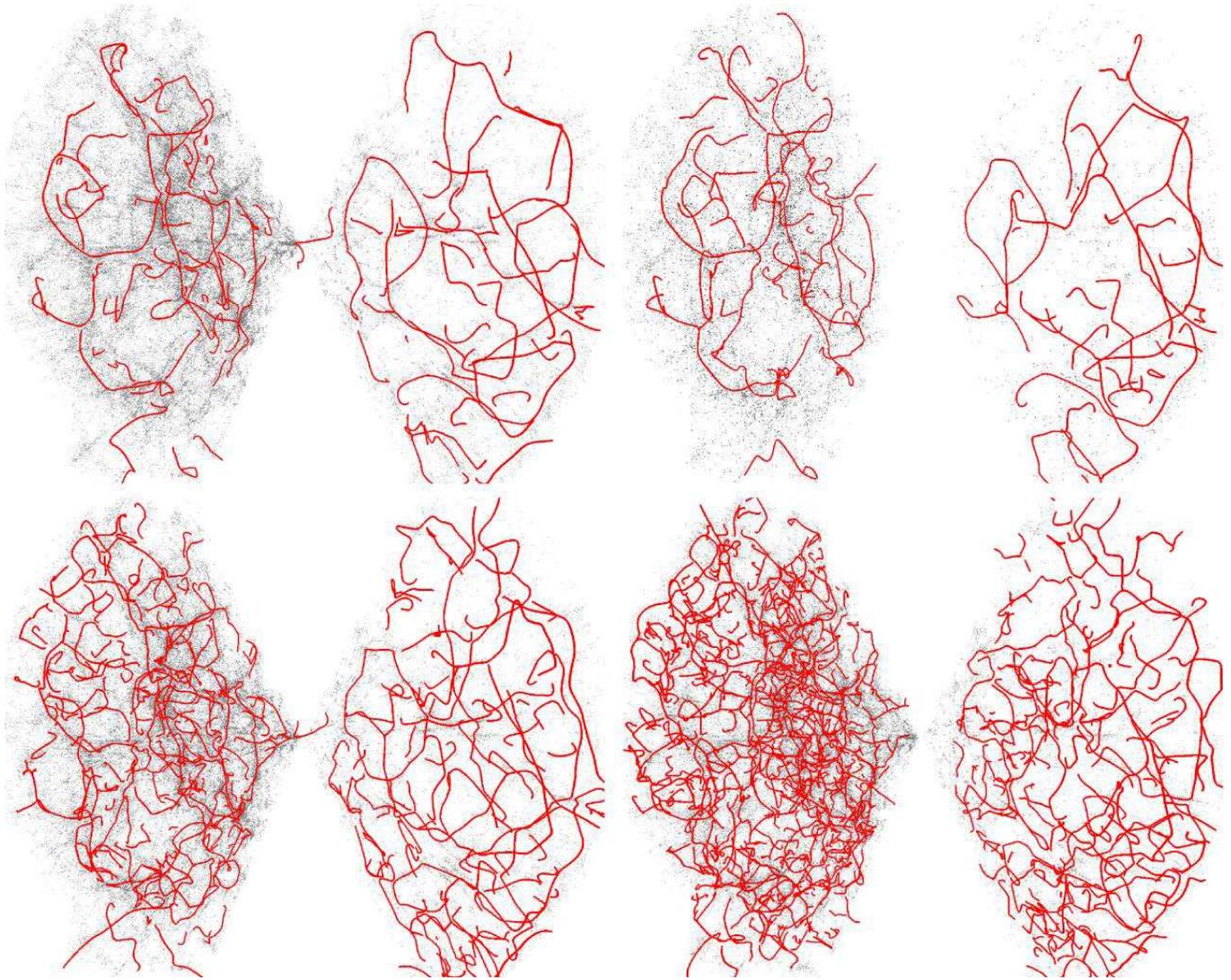} \end{center}
\caption{The skeleton of the SDSS compared to the corresponding galaxy distribution.
{\em Top left and right panels:} the galaxy distribution, respectively in DR4-350 and its volume-limited counterpart,
DR4-VL350 with the superimposed skeleton measured for a smoothing scale $L=16.4$ Mpc. {\em Bottom left and right panels:} the skeleton measured in DR4-350, for $L=10.9$ and $8.2$ Mpc, respectively.  
To have a better feeling of what the results look like in 3D,
a mpeg movie is  also available in the electronic edition of the {\it
Astrophysical Journal}, where the skeleton is measured on the full SSDS-DR4 data for $L=16.4$ Mpc.
 \label{fig2}}
\end{figure*}

Figure~\ref{fig2} shows the skeleton measured in DR4-350 for 3 smoothing scales
$L=16.4$, $10.9$ and $8.2$ Mpc, {\em (top left and 2 bottom panels) }
while the measurements of its length, ${\cal L}$
as a function of $L$ are summarized in Table~\ref{tab1}. 

As expected, the skeleton matches the intuitive visual definition of what a filament is, and
its length and complexity increase with the inverse of $L$.
Notice on Fig.~\ref{fig2} that the prominent features of the skeleton remain mostly independent
of smoothing: decreasing $L$ essentially adds new branches to the skeleton, corresponding to
finer structures in the galaxy distribution. In other words, the skeleton grows like a tree,
while $L$ decreases. The overall scale dependence of the measured length,
${\cal{L}} = (52500\pm6500) (L/{\rm Mpc})^{-1.75\pm0.06}\rm{Mpc}/(100\,\rm{Mpc})^{3}$, is in good 
qualitative agreement with the expected trivial power-law in the scale-free case, ${\cal L} \propto L^{-2}$
(SPCN). 

These results match very well the predictions of the standard $\Lambda$CDM 
model (compare MOCK to DR4-350). This allows one to use the mock catalogs as a solid baseline
to test possible observational and dynamical effects on the skeleton, as discussed now,
 using Table~\ref{tab1} as a guideline. 
{\em Incompleteness and discreteness} effects can simultaneously be tested by comparing DR4-350 
to its volume-limited counterpart, DR4-VL350, which probes only 15 percent of the galaxies 
available in DR4-350 (see the top panels of Fig.~\ref{fig2}). 
They have little impact on the skeleton, changing its length
by at most 3 percent. {\em Edge effects} arise from the
particular geometry of the SDSS. They can be probed by comparing MOCK to   MOCK-AS. 
They have a small but systematic impact on the measured length of the skeleton, which is increasingly
overestimated with scale, from about 1 percent for $L=8.2$ Mpc to 8 percent for $L=16.4$ Mpc.
In terms of scaling behavior, ${\cal L} \propto L^{-\alpha}$, $\alpha$ is therefore slightly 
underestimated, which explains partly the slight deviation from the expectation $\alpha=2$,
in addition to the scale dependence of the power-spectrum of the density fluctuations.
{\em Cosmic variance} should be small: when estimated from the dispersion among 25 realizations
of MOCK, it increases with smoothing scale, as expected,
from a 2 percent error for $L=8.2$ and $L=10.9$ to a $5$ percent error for $L=16.4$.
{\em Redshift distortion} effects, discussed at length in SPCN,
can be tested by comparing MOCK to MOCK-PS.
They have negligible impact on the measurements, well within the cosmic variance. 
Finally, since the skeleton probes overdense regions of the universe and 
large smoothing scales are considered, the measurements are expected to be rather insensitive to effects of {\em biasing}
and to be dominated by the predictions of  {\em linear} dynamics. This has been fully confirmed when comparing the skeleton of simulations at $z=0$ and in the initial conditions, at least
in the $\Lambda$CDM cosmogony framework. Moreover, the small amplitude of the change of length of the skeleton when nonlinear biasing is applied can be checked in Table~\ref{tab1} (compare MOCK-NB and MOCK).

\section{Discussion: 
 a test of LSS formation models}
\label{sec:discu}

\begin{figure}[h]
\begin{minipage}[b]{0.95 \linewidth}
\centering \includegraphics[width=8cm]{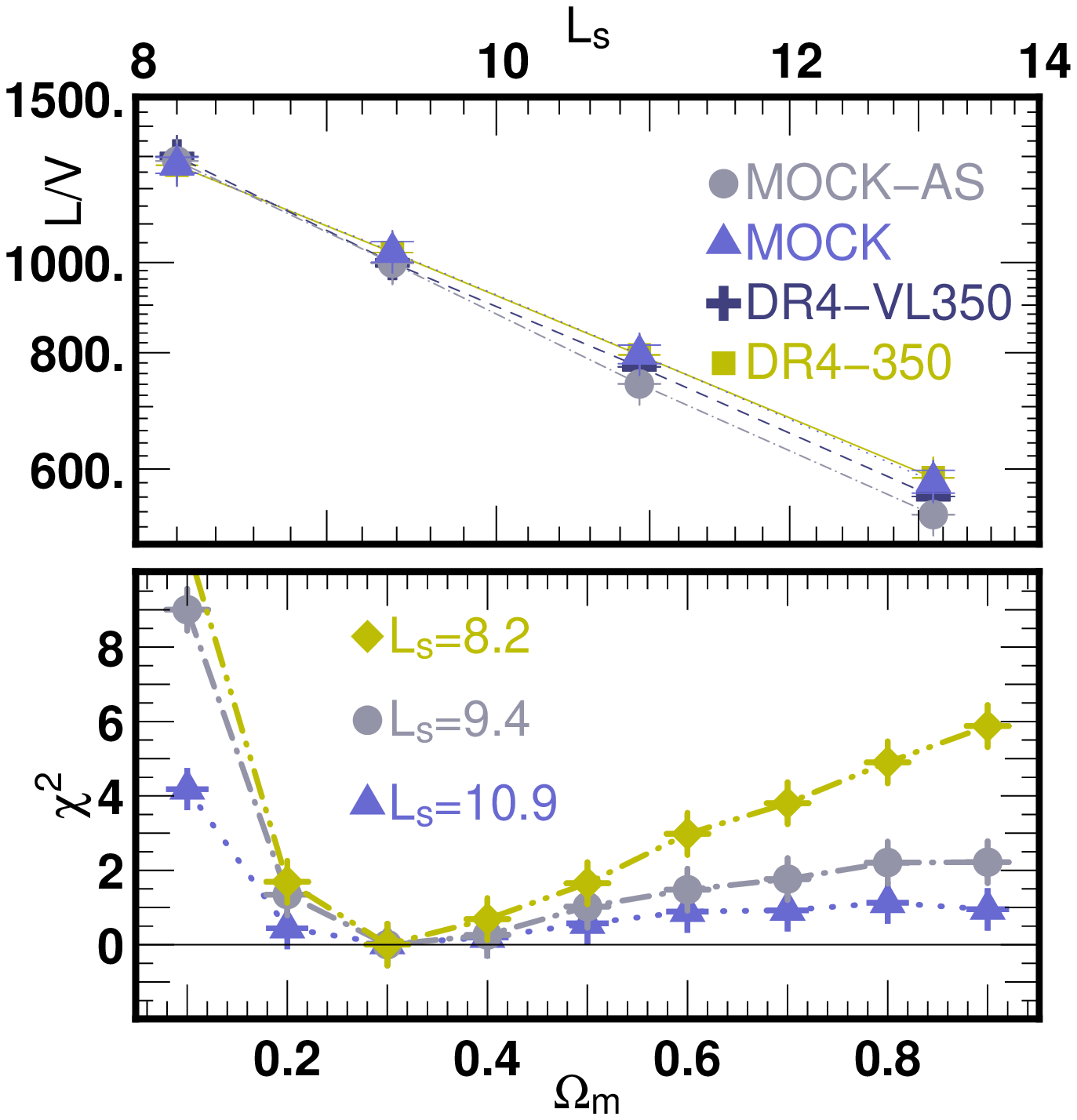}
\caption{{\em Top panel:} skeleton length per unit volume (in Mpc$/$($100$ Mpc)${}^3$) as a function of
  smoothing length for mock catalogs and SDSS showing the very good agreement (as explained in the text).
  The error bars correspond to the cosmic variance, which is estimated via
  25 realizations of the mock.
  {\em Bottom panel:} $\chi^2$ corresponding to
  the squared difference between the  total length per unit volume
  of the mock catalog and that of the SDSS, in units of the RMS of the mock;
  These $\chi^2$ curves
  yield a confidence interval for $\Omega_{\rm matter}$ (for a flat universe)
  of $[0.25,0.4]$ at one-$\sigma$ level
  and $[0.15,0.75]$ at two-$\sigma$ level.
  The simulations are $256^3$  dark matter particles of a given $H_0=70$ km/s/Mpc
  using an Eke prescription \citep[]{Eke96} for the normalization of the spectrum.
\label{fig3}}
\end{minipage}
\end{figure}

In this letter a method to probe the filamentary structure 
in the galaxy distribution, involving the
extraction of the {\em local skeleton} from the data and measuring 
its length per unit volume, ${\cal L}$, was tested on the SDSS and mock
catalogs. The length of the skeleton was found to be a robust statistic
in the scaling regime $8.2 \leq L \leq 16.4$ Mpc, rather insensitive
to nonlinearities,  biasing, redshift distortion, incompleteness and cosmic variance.
The results were however slightly affected by edge effects due to the geometry of the SDSS (see Table 1 Column 6). 
Still, one observes an excellent agreement with the $\Lambda$CDM concordant
model. (See figure 3, top panel).

One question remains: is the length of the skeleton a discriminant measure of large scale
structure? In theory, the answer is positive: for a  Gaussian  field, ${\cal L}$ 
depends on the shape of the power-spectrum of density
fluctuations, $P(k)$, through its moments of order $2m$, $\int k^{2m+2} P(k) \exp(-k^2L^2) {\rm d} k$, up to $m=3$,
leading to the approximate scaling ${\cal L} \propto (6.2+n) L^{-2}$ for $P(k) \propto k^n$ (SPCN).
To demonstrate that this spectral dependence can be used to constrain models of large scale structure in practice, nine flat universe simulations 
were carried out with GADGET-2,  involving
$256^3$ particles and with the same cosmological parameters as previously used except that 
$\Omega_{\rm matter}$ was left as a free variable in the range $0.1 \leq \Omega_{\rm matter} \leq 0.9$.
From each of the simulations, 25 mock catalogs were extracted,
in which ${\cal L}$ was estimated. These measurements were used to perform standard 
$\chi^2$ analysis to find the best matching value of $\Omega_{\rm matter}$ 
for the SDSS, using MOCK as reference.
The final $1\sigma$ constraint is $0.25 < \Omega_{\rm matter} < 0.4$
(See figure 3).

This clearly demonstrates that the length of the skeleton is a discriminant estimator,

which might prove to be a real alternative
to traditional two-point statistics estimators which are extremely sensitive to the bias in the
nonlinear stage of gravitational instability. 
 The local skeleton extraction also
 opens   new paths of investigation for  the structure analysis of
galactic or  dark matter distribution,  with the prospect  of defining
quantitatively the  locus of filaments. In particular, it  will allow
astronomers to  carry measurements (velocity, pressure...) along the
main motorways of galactic infall \citep[]{Aubert04}.

\vskip 0.5 cm
\section*{ ACKNOWLEDGMENTS}
This  work  was  carried  out within  the  Horizon  project,
\texttt{www.projet-horizon.fr}. We thank  the  SDSS  collaboration,
 \texttt{www.sdss.org}, for publicly releasing the  DR4 data. 
The computational means used to perform the $512^3$ simulation (IBM POWER4) 
were made available to us by IDRIS. We thank S. Prunet, D. Aubert, J. Devriendt and D. Pogosyan for  useful comments.




\end{document}